\documentclass[%
linenumbers,
10,
jpb,
]{iopart}
\usepackage{graphicx}% Include figure files
\usepackage{hyperref}
\usepackage{here}
\usepackage{indentfirst}
\usepackage{dcolumn}% eqnarray table columns on decimal point
\usepackage{bm}% bold math
\usepackage{braket}
\usepackage{hyperref}
\usepackage{txfonts}
\usepackage{ulem}
\usepackage{color}
\usepackage{url}
\newcommand{\ueda}[1]{\textcolor{black}{#1}}
\newcommand{\ota}[1]{\textcolor{black}{#1}}
\newcommand{\yamazaki}[1]{\textcolor{black}{#1}}
\newcommand{\hatada}[1]{\color{black}{#1}}
\newcommand{\didier}[1]{\textcolor{black}{#1}}
\newcommand{\otarev}[1]{\textcolor{black}{#1}}
\newcommand{\hatadarev}[1]{\textcolor{black}{#1}}

\begin{document}
\title[{Extracting bond length variation of dissociating CO$^{2+}$ from forward and backward intensities}]{Theory on polarization-averaged core-level molecular-frame photoelectron angular distributions: II. \yamazaki{Extracting the X-ray induced fragmentation dynamics of carbon monoxide dication from forward and backward intensities}}
\author{
F Ota$^1$, 
K Hatada$^2$,
D S\'ebilleau$^3$, 
K Ueda$^4$ and 
K Yamazaki$^{5,6}$}

\address{$^1$ Graduate School of Science and Engineering for Education, University of Toyama, Gofuku 3190, Toyama 930-8555, Japan}
\address{$^2$ Faculty of Science, Academic Assembly, University of Toyama, Gofuku 3190, Toyama 930-8555, Japan}
\address{$^3$ D\'epartement Mat\'eriaux Nanosciences, Institut de Physique de Rennes, UMR UR1-CNRS 6251, Universit\'e de Rennes, F-35000 Rennes, France}
\address{$^4$ Institute of Multidisciplinary Research for Advanced Materials, Tohoku University, Katahira 2-1-1, Aoba-ku, Sendai 980-8577, Japan}
\address{$^5$ Institute for Materials Research, Tohoku University, 2-1-1 Katahira, Aoba-ku, Sendai 980-8577, Japan}
\address{$^6$ Attosecond Science Research Team, Extreme Photonics Research Group, RIKEN Center for Advanced Photonics, RIKEN, 2-1 Hirosawa, Wako, Saitama, 351-0198, Japan.}

%\author{Fukiko Ota}
%\address{Graduate School of Science and Engineering for Education, University of Toyama, Gofuku 3190, Toyama 930-8555, Japan}
%\author{Keisuke Hatada}
%\address{Faculty of Science, Academic Assembly, University of Toyama, Gofuku 3190, Toyama 930-8555, Japan}\author{Didier S\'ebilleau}
%\address{D\'epartement Mat\'eriaux Nanosciences, Institut de Physique de Rennes, UMR UR1-CNRS 6251, Universit\'e de Rennes, F-35000 Rennes, France}
%\author{Kiyoshi Ueda}
%\address{Institute of Multidisciplinary Research for Advanced Materials, Tohoku University, Katahira 2-1-1, Aoba-ku, Sendai 980-8577, Japan}
%\author{Kaoru Yamazaki\footnote{Present address: Attosecond Science Research Team, Extreme Photonics Research Group, RIKEN Center for Advanced Photonics, RIKEN, 2-1 Hirosawa, Wako, Saitama, 351-0198, Japan.}}
%\address{Institute for Materials Research, Tohoku University, 2-1-1 Katahira, Aoba-ku, Sendai 980-8577, Japan }
% \affiliation{Boston University, Department of Physics, Boston, MA 02215}

\eads{\mailto{hatada@sci.u-toyama.ac.jp}, \mailto{kaoru.yamazaki@riken.jp}}
\date{\today}

\begin{abstract}
% We demonstrate series of calculations for polarization-averaged molecular-frame photoelectron angular distributions (PA-MFPADs) emitted from $1s$ orbital of oxygen atom of dicationic carbon monoxide excited by irradiating two color X-ray free electron laser pulses.
% In this study, the calculations are performed in terms of the Full-potential multiple scattering theory, and the potentials of each scattering site are calculated using Multi-Configuration Self Consistent Field (MC-SCF) method in order to take the electronic holes into account. We find the remarkable result that the PA-MFPADs are very sensitive to the molecular geometry while the influence of electronic structure does not play an important role.
%
%Visualizing structural changes of molecules in reactions in real time has been a long-standing dream in chemical physics, physical chemistry, and biochemistry. 
Recent developments in high repetition-rate X-ray free electron lasers (XFELs) such as the European XFEL and the LSCS-II, combined with coincidence measurements \hatada{at} the COLTRIMS--Reaction Microscope, is now opening a door to realize the long-standing dream to create molecular movies of photo-induced chemical reactions in gas-phase molecules. 
In this paper, we propose a new theoretical method to experimentally visualize the dissociation of diatomic molecules via time-resolved polarization-averaged molecular-frame photoelectron angular distributions (PA-MFPADs) measurements using the COLTRIMS--Reaction Microscope and the two-color XFEL pump-probe set-up. We used first and second order scattering theory within the Muffin-tin approximation\hatadarev{, which is valid for a sufficiently high kinetic energy of photoelectron, typically  above 100 eV, and for long bond lengths}. This leads to a simple EXAFS-type formula for the forward and backward scattering peaks in the PA-MFPADs structure. This formula relies only on three semi-empirical parameters obtainable from the time-resolved measurements. It can be used as a "bond length ruler" on experimental results.
The accuracy and applicability of the new ruler equation are numerically examined against \yamazaki{the PA-MFPADs of CO${}^{2+}$ calculated with Full-potential multiple scattering theory as a function of the C-O bond length reported in the preceding work~\cite{Ota2020a}.} %We computed the PA-MFPADs using Full-potential multiple scattering theory \ueda{reported in the preceding paper.} %scattering
%combined with electrostatic potential calculated using ab-initio \comment { second order } multirefenence perturbation theory. 
\ueda{The bond lengths retrieved from the the PA-MFPADs via our EXAFS-like formula coincide \yamazaki{within an accuracy of 0.1 \r{A}} with the original C-O bond lengths used in the reference \textit{ab-initio} PA-MFPADs.}
%The bond lengths retrieved from the PA-MFPADs via the Young's formulas also reproduce the original C-O bond lengths used in the reference \textit{ab-initio} PA-MFPADs within the relative error of \yamazaki{6-11 \% }.
%
We expect time-resolved PA-MFPADs to become a new attractive tool to make molecular movies visualizing intramolecular reactions.
\end{abstract}

\noindent{\it Keywords\/}: \hatada{PA-MFPADs, Multiples Scattering theory, Coincidence measurement, FEL}

\submitto{\jpb}

\maketitle
\ioptwocol

\section{Introduction}

Visualizing structural changes of molecules in reactions in real time has been a long-standing dream in chemical physics, physical chemistry, and biochemistry~\cite{
%Zweil2008, I couldn't find this
Zewail2014, 
Miller2014}.
\hatada{In particular, seeing structural changes of photoexcited free molecules is of fundamental interest but it is still a challenge.}
%Visualizing a structural change of a photoexcited free molecule is of fundamental interest but still a challenge.
The advent of X-ray free electron lasers (XFEL)  delivering extremely short X-ray pulses down to a few femtoseconds~\cite{Emma2010} and of a mega-electron-volt ultrafast electron diffraction (UED) system~\cite{Weathersby2015} at SLAC opens up new pathways to \hatada{investigate these} structural changes
%of the photoexcited molecules
in real time~\cite{Minitti2015,Wolf2019}. 

As an alternative approach, time-resolved photoelectron diffraction (PED) employing an XFEL as a photoionizing source was recently  proposed~\cite{Krasniqi2010, Kazama2013}. PED is a technique well-established as an analytic tool in surface science~\cite{Woodruff2008}, where the directions of core-level photoelectrons emitted from a specific site are measured with respect to the sample orientation. In the gas-phase PED experiments, the molecular axis needs to be fixed in space. In case of synchrotron radiation experiments, this may be realized by detecting fragment ions in a momentum-resolved manner and core-level photoelectrons are then measured in
coincidence with these fragment ions, using a COLTRIMS--Reaction
Microscope~\cite{Ullrich2003}. These measurements provide us with photoelectron angular distributions in molecular frame (MFPADs), which are equivalent to PED. Indeed, extractions of the three dimensional structure defined by the direction of the bonds~\cite{Williams2012} and the bond length~\cite{Fukuzawa2019} from the measured polarization-averaged MFPADs (PA-MFPADs) have been demonstrated.

Some attempts towards time-resolved MFPADs measurements with XFELs were reported~\cite{Rouzee2013, Boll2013, Nakajima2015, Minemoto2016} but there has been no report on time-resolved studies of molecular structural changes so far, due to complexity of the time-resolved MFPADs measurements with XFELs. Coincidence experiments to measure MFPADs using synchrotron radiation as a light source are well established,  but up to now, they were practically impossible with a low repetition-rate XFELs as three different lasers, one for actively arraying the molecule in space, one for exciting the molecule to initiate a reaction, and the XFEL as a light source for the MFPADs measurements, had to be operated in a synchronized manner~\cite{Rouzee2013} in order to perform the experiment. 

The first high repetition-rate XFEL, the European XFEL \cite{Tschentscher2017}, has just started operating for users. Its high repetition-rate opens up the door to coincidence experiments with the COLTRIMS--Reaction Microscope routinely used for MFPADs measurements, and Kastirke {\it et al.}~\cite{Kastirke2020} have just reported the first successful implementation of this technique at the soft X-ray beam line of the European XFEL. In the USA, LCLS-II will soon start operating with a high repetition-rate in the soft to tender X-ray regions~\cite{Duris2020}. Furthermore, two-color XFEL operations will be available at the European XFEL~\cite{Serkez2020} and at the LCLS-II~\cite{Duris2020}, allowing to perform pump-probe experiment for multiple edges scheme such as C(1s) pump O(1s) probe etc. These two-color operations combined with the COLTRIMS--Reaction Microscope will provide the scientific community with another attractive pathway to make molecular movies to visualize the intramolecular reaction occurring, e.g., in dicationic states. 
%Noting these advances of the light sources in mind, we have started simulation studies for MFPADs of molecules that go structural changes. We report here the first such study on CO${}^{2+}$.

In the present work, we consider the process where the first XFEL pulse produces CO${}^{2+}$ via the Auger decay that follows the core ionization of CO, and the second XFEL pulse kicks out an oxygen $1s$ electron from CO${}^{2+}$. Using \ueda{first and second order Muffin-tin multiple scattering theory}, we derive analytical EXAFS-type expressions for the backward-forward scattering peaks. 
%double-slit-interference-type expression for the angles where PA-MFPADs exhibit peaks corresponding to double-slit interference
 Employing the Full-potential multiple scattering PA-MFPADs data reported in the preceding paper~\cite{Ota2020a}, we introduce a parameterized formula to describe the EXAFS-like oscillations in the forward-backward \hatada{intensity} ratio. This formula relies only on three semi-empirical parameters that account for the effects of Full-potential multiple scattering on the PA-MFPADs.  Because of it simplicity, it can be used to extract the bond length from the experimental PA-MFPADs, which varies as a function of pump-probe delay.

 % =================================================================== 
%It is revealed that structural information of molecule can be extracted from the polarization-averaged (PA) MFPADs~\cite{Williams2012, Fukuzawa2019, Kastirke2020}.
%In this section, with two-color XFEL pump-probe scheme in mind, we present our computational results of oxygen $1s$ PA-MFPADs of CO${}^{2+}$ as a function of C-O bond length.
%In the two-color XFEL pump-probe scheme, the pump pulse removes an electron from carbon $1s$ orbital, then this eventually induces C ($1s$) $KVV$ Auger decay and produces CO${}^{2+}$ in various electronic excited states. Subsequent probe pulse excites futher one more electron from oxygen $1s$ orbital to continuum state as a photoelectron, then its diffraction pattern is observed as PA-MFPADs as a function of pump-probe delay time or C-O bond length.
%
%\section{Analytical formulas of PA-MFPADs for diatoms}
\section{\hatada{Analytical Expression for PA-MFPADs to Study Dynamics of Dissociating CO\textsuperscript{2+}}}
\hatada{In this section, we derive analytical formula\didier{s} to extract the C-O bond length during the dissociation process. We first calculate the PA-MFPADs within our Full-potential multiple scattering theory in order to monitor the variations in the PA-MFPADs patterns with the changing C-O bond length. The details of the theory are found in the preceding paper~\cite{Ota2020a}. Then, in Section~\ref{sec:forwardbackward}, we derive analytical expressions for the ratio of the backward-intensity to forward-intensity s a function of the C-O bond length, using the single-scattering Plane Wave approximation within \didier{the} Muffin-tin approximation.}

%\subsection{Numerical results of PA-MFPADs for dissociating CO\textsuperscript{2+}}
\hatada{\subsection{Numerical Results of PA-MFPADs with Full-potential Multiple Scattering Theory for Dissociating CO\textsuperscript{2+}}}
%
% ======================================================================
\begin{figure*}[htb]
\includegraphics[width=\linewidth]{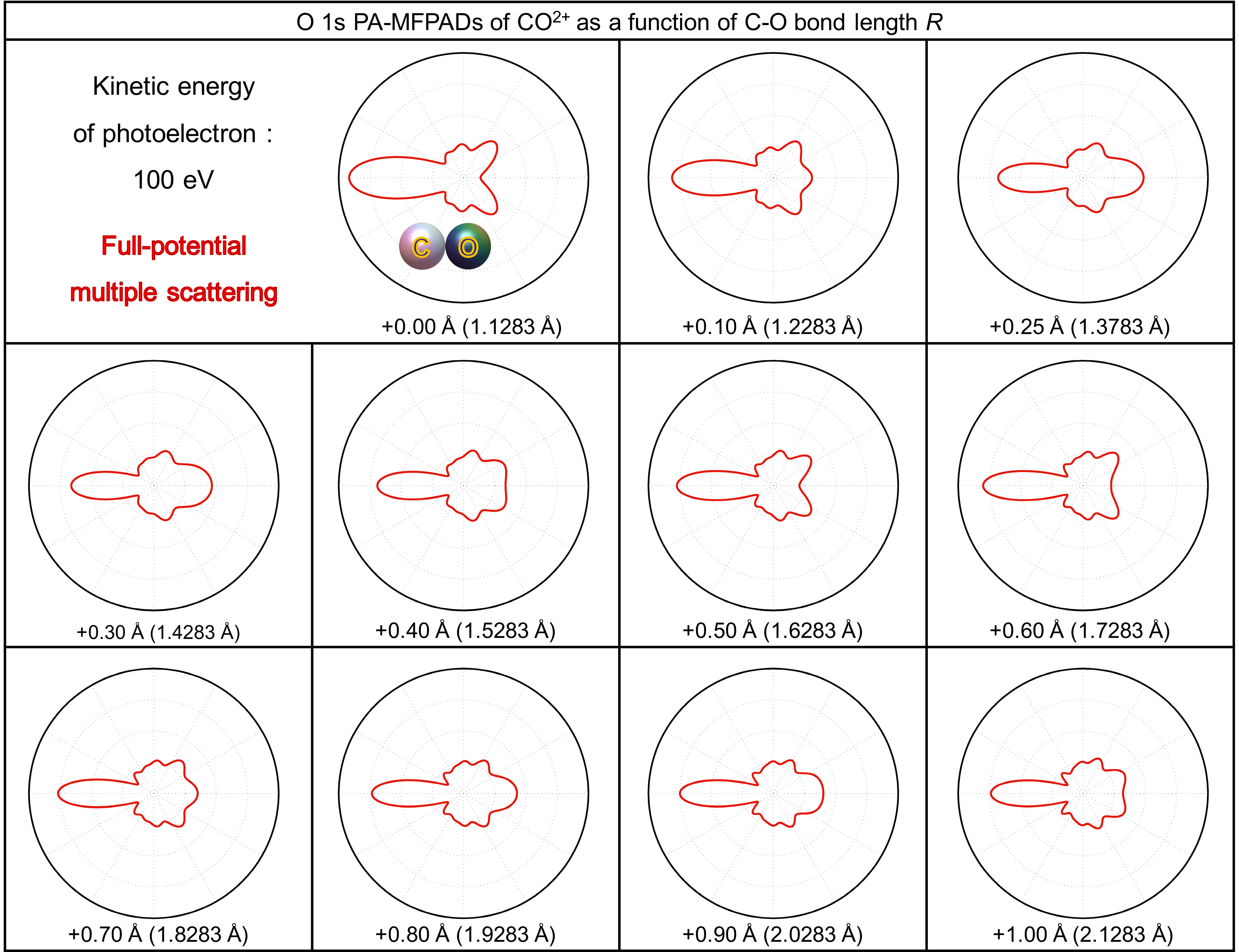}
\caption{\ota{The oxygen $1s$ }\yamazaki{PA-MFPADs of CO${}^{2+}$ calculated with the Full-potential multiple scattering method as a function of the C-O bond length $R$ in \AA. Here, the photoelectron energy is \ota{$100$ eV}. More details are provided in  the preceding paper \cite{Ota2020a}.}
\label{fig:peaks}
}
\end{figure*}
% ======================================================================
% ======================================================================
%\begin{figure*}[htb]
%\includegraphics[width=\linewidth]{fig%ure/all.png}
%\caption{Newer version of figure1. $V_0$ %is set as 0 for Muffin-tin and %Full-potential. The basis set used is %6-31G* (something like this) for quick %check.}
%\label{fig:peaks_new} 
%}
%nd{figure*}
% ======================================================================

% ===============================================================%=======
%\begin{figure*}[ht]
%\includegraphics[width=\linewidth]{MT1_FP.png}
%\caption{\yamazaki{PA-MFPADs and its peak position of CO${}^{2+}$ calculated by the Full-potential multiple scattering calculations as a function of C-O bond length. We choose photoelectron energy $E = 100 $ eV. For the detail, see the preceding paper.}
%\ota{Muffin-tin single-scattering calculations are indicated with green line. The peak positions of Muffin-tin single-scattering calculation and Full-potential multiple scattering calculation are close, while its intensities are different.}
\label{fig:MT1_FP} 
%}
%\end{figure*}
% ======================================================================
%
% ======================================================================
%\begin{figure}[htb]
%\includegraphics[width=1.0\linewidth]{oscillation_ratio_solo.png}
%\caption{
%%\label{fig:oscillation}
%Newer figure of figure2. $V_0$ is set to the FP's one. The basis set is ANO.
%}
%\end{figure}

% ===============================================================%=======

\hatada{In order to compute the PA-MFPADs for a photoelectron of momentum $\textbf{k}$}, we use the formula
\begin{eqnarray}
\label{eq:pamfpads}
\left<I({\bf k})\right>_{\varepsilon}
&=
%frac{8 \pi^2 \alpha \hbar \omega}{3}
%\sum_{m_c}\sum_{n=1}^3
%\left| \,
% \displaystyle{
% \sum_{L}
% B_{1n}^{\,i*} ({\bf k}) \,
% M_{L_c,1n} (\hat{\bf \varepsilon})
% }
% \right|^2,
 \frac{8 \pi^2 \alpha \hbar \omega}{3}
 \sum_{n=-1}^1 \sum_{m_c}
 \Bigg|
 \sqrt{
 \frac{4\pi}{3}
 }
 \sum_{LL'}
 B_L^{\,i *} ({\bf k})
 \nonumber \\
 &
 \hspace{0.3cm}
 \times
 C(L,1n,L_c) \int dr \,
 r^3 \,
 R_{L'L} (r;k) \,
 R_{L_c}^{\,c} (r) \,
\Bigg|^2 ,
\end{eqnarray}
\hatada{where}
\begin{eqnarray}
B_L^{\,i *} ({\bf k})=&=&
 \sum_{jL'}
 \tau_{LL'}^{ij} \,
 I_{L'}^{\,j} ({\bf k}),
\end{eqnarray}
\hatada{and}
\begin{eqnarray}
\tau &\equiv &\left( T^{-1} - G \right)^{-1}=
T \left( 1 - G T \right)^{-1}, \label{eq:tau}
\\
 I_{L}^{\,i} ({\bf k}) &\equiv& %& i^{\,l}
 \sqrt{\frac{k}{\pi}}\, 
 e^{i{\bf k }\cdot{\bf R}_{io}}
 \mathcal{ Y }_{L}(\hat{\bf k }).
\end{eqnarray}
$i$ and $j$ identify the scattering sites.
$\alpha$ is the fine structure constant, $\textbf{R}_{io}$ is the vector connecting the origin to the center of scattering site $i$ and the $\mathcal{ Y }_{L}$ are the real spherical harmonics with angular momentum $L=(l,m)$. In the second line of equation~\ref{eq:pamfpads}$, C(L,1n,L_c)$ is a Gaunt coefficient with real spherical harmonics and the integral is the radial integral between the local solution and the core wave function. In equation~\ref{eq:tau}, $T$ is the transition operator (or scattering $T$ matrix) and $G$ is the KKR structure factor~\cite{Korringa1947,Kohn1954}.
More details are reported in our preceding paper~\cite{Ota2020a}.

Figure~%\ota{\ref{fig:MT1_FP}} 
\ueda{\ref{fig:peaks} depicts} the PA-MFPADs calculated as a function of  the \hatada{C-O bond length} $R$. The calculations have been performed  \hatada{with our Full-potential multiple scattering method. The electron charge density was computed \yamazaki{at the RASPT2/ANO-RCC-VQZP level of theory as implemented in MOLCAS 8.2~\cite{Aquilante2016}} for the major excited state 1$\sigma^{-1}$5$\sigma^{-2}$ (i.e. O 1$s^{-1}$ HOMO$^{-2}$) of the Auger final state~\cite{Cederbaum1991}.}
The kinetic energy of photoelectron was set to 100 eV, where energies are counted from the interstitial potential $V_0$, which itself is defined as the average of the potential of the scattering cells. 
\hatada{We note that the forward-intensity peak towards the carbon atom has a large lobe due to the focusing effect.}
The backward-intensity peak lobe \hatada{in the opposite direction} oscillates as a function of $R$. These oscillations correspond to the EXAFS oscillations. 
%As seen in the subsequent subsections, one period of the oscillation corresponds to the increase in $R$ by $\Delta R =\pi/k$ within the framework of a single-scattering and the Plane Wave approximations, where $k$ is the photoelectron momentum in atomic units. 
%There are several small lobes between the forward- and backward-\hatada{intensity} peaks. These lobes move from the backward to forward with the increase in $R$ and increase the number by one for one period of oscillation of the backward-intensity peak. 
%In a single-scattering and the Plane Wave approximations, labelling angles $\theta_1$ and $\theta_2$ for the last two lobs (i.e. two closest to the backward-intensity peak) relative to the forward-scattered peak (see figure~\ref{fig:geometry}), we have $R=2\pi/k(\cos\theta_2-\cos\theta_1)$. See Sec.~\ref{sec:flowershape} for further details. 
Therefore, the shape of the PA-MFPADs contains certainly the information about the bond length. We will hereafter derive analytical EXAFS-like formulas that can be used to extract the \hatada{C-O} bond length from the PA-MFPADs. 
\\
%It shows that the PA-MFPADs structure
%in the "Backward" direction ( direction of oxygen seen from carbon,
%indicated with blue arrow in figure~\ref{fig:bondlength_dependency} )
%is sensitively changed depending on C-O bond length,
%while that in the "Forward" direction (indicated with red arrow) is not affected much.

% =================================================================
%\section{Extracting the structural dynamics from PA-MFPADs}
%In the previous section, we saw that the PA-MFPADs have correspondence with C-O bond length $R$, and we confirmed our Full-potential multiple scattering treatment is very good tool to predict PA-MFPADs. In this section, we analytically derive simple expressions for PA-MFPADs intensity as a function of $k$ and $R$ for heteronuclear diatomic molecules within the Muffin-tin and the Plane Wave approximations, taking single- or double-scatterings into account.
% ======================================================================

%subsection{The first and second order Muffin-tin approximations}
\subsection{\hatada{Single and Double Scattering Approximations with the Muffin-tin Approximation}}

In this subsection, we derive the PA-MFPADs intensity
within \didier{the} Muffin-tin approximation, in which \didier{the} electrostatic potential and the charge density are spherically averaged on each \hatada{scattering site}, thereby neglecting their anisotropies. This will allow us in the following subsections to analytically study the relationship between the  PA-MFPADs patterns and the \hatada{C-O} bond length.

\hatada{In the framework of \didier{the} Muffin-tin approximation,}
the radial part of local solution $R_{LL'}(r_i;k)$ and the transition matrix $T_{LL'}^{\,i}$ are diagonal in angular momentum,
\begin{eqnarray}
 R_{LL'}(r_i;k)
 \rightarrow %\mbox{Muffin-tin~approx.} &
 %\xrightarrow[ \mbox{Muffin-tin~approx.} ]{}&
 R_{l}(r_i;k) \,
 \delta_{LL'}
 \\
 T^{\,i}_{LL'}
 % & =
 % -\sum_{L''}S^{\,i}_{LL''}\,(E^{\,i})_{L''L'}^{-1}
 % \\
 \rightarrow %\mbox{Muffin-tin~approx.} &
 %\xrightarrow[ \mbox{Muffin-tin~approx.} ]{}&
 % -\sum_{L''}S^{\,i}_{l}\, \delta_{LL''} \,(E^{\,i})_{l'}^{-1} \, \delta_{L''L'}
 % =
 % -S^{\,i}_{l} \,(E^{\,i})_{l}^{-1} \, \delta_{LL'}
 % \equiv
 T_{l}^{\,i} \, \delta_{LL'} 
\end{eqnarray}
Consequently, the intensity of the PA-MFPADs becomes
\begin{eqnarray}
\left<I({\bf k})\right>_{\varepsilon}
&\sim
 \frac{8 \pi^2 \alpha \hbar \omega}{3}
 \sum_{n=-1}^1 
 \sum_{m_c} 
 \Bigg|
 \sqrt{
 \frac{4\pi}{3}
 }
 \sum_{L}
 B_L^{\,i *} ({\bf k})
 \nonumber \\
 &
 \hspace{0.3cm}
 \times
 \otarev{ C(L,1n,L_c)\, }
 \int dr \,
 r^3 \,
 R_{l} (r;k) \,
 R_{L_c}^{\,c} (r)
\Bigg|^2 
 \nonumber \\
 &\hspace{-1cm}=
 \frac{8 \pi^2 \alpha \hbar \omega}{3}
 \sum_{n=-1}^1 
 \sum_{m_c} 
 \Bigg|
 \sqrt{4\pi}
 \sum_{L}
 B_L^{\,i *} ({\bf k}) \,
 \otarev{ C(L,1n,L_c)\, }
 M_{L_c}^{\,l}
\Bigg|^2 
\end{eqnarray}
where $
M_{L_c}^{\,l}
\equiv
\sqrt{1/3}
%(1/3)^{1/2}
\int dr \,
 r^3 \,
 R_{l} (r;k) \,
 R_{L_c}^{\,c} (r) \,
 $ is \hatada{the transition matrix} describing the excitation of the photoelectron.
 %and $\left< ... \right>_{\varepsilon}$ represents that the polarization average was taken over polarization of the photon. We used the notation, $L=(l,m)$, for angular momentum
We specialize now in the photoemission from an oxygen $1s$ \hatada{core} orbital, $L_c=(0,0)$, so that
% \comment { For the photoemission process from $1s$ core orbital, , } it becomes simpler, 
%\comment {For simplicity, we consider only the $s$-partial wave ($L_c=(0,0)$) for the photoemission process from $1s$ core orbital}
\begin{eqnarray}
\left<I({\bf k})\right>_{\varepsilon} &\sim
 \frac{8 \pi^2 \alpha \hbar \omega}{3}
 \left|
 M_{00}^{\,1} 
\right|^2 
 \sum_{n=-1}^1 
 \left|
 B_{1n}^{\,i} ({\bf k})
\right|^2 .
\end{eqnarray}
 % ----------------------------------------------------------------------
% ===================================================================
For a sufficiently high kinetic energy of photoelectron and a long bond length\hatada{, namely a large value of $kR$,}
 %\yamazaki{(large $kR$)}
the spectral radius of matrix $GT$ in equation~\ref{eq:tau} becomes less than one\hatada{, i.e. $\rho(GT)<1$}~\cite{Sebilleau2012}.
\hatada{Under this condition, we can expand the multiple scattering  matrix $\tau$  into a series expansion} 
,
\begin{eqnarray}
 \tau = T \left( 1 - G T \right)^{-1} \approx & T + TGT + TGTGT + \cdots. 
\end{eqnarray}
The first term corresponds to the direct photoemission process 
on the absorbing \hatada{atom} and the second term includes all the scattering paths contributing to the single-scattering process. In the same way the higher terms describe the higher scattering orders.
% If this expansion truncated at second term, 
% it corresponds that the calculation is performed with including the scattering effect up to single-scattering.
 % ----------------------------------------------------------------------
 \par
 
In order to help the understanding of the PA-MFPADs, we introduce a 
further simplification into our equations, replacing the KKR structure factors $G_{LL'}^{ij}$ by their plane wave approximation \cite{Rehr1986},
% 
% G_{LL'}^{ij} \equiv& 4 \pi \sum_{L''} \,
%                  i^{l+l''-l'} \, C (L''L'L) \,
%                   \tilde { H }^{+}_{L''} ({\bf R}_{ij}, k ) \\
%
\begin{eqnarray}
 G_{LL'}^{ij}
  \sim& 
  - 4 \pi \,
  i^{\, l-l'} \, 	       
  \frac{\,e^{ikR_{ij}}\,}{\,R_{ij}\,}	  
  \mathcal{ Y }_{L}   ({\bf \hat {R} } _{ij}) \,
  \mathcal{ Y }_{L'}    ({\bf \hat {R} } _{ij}), \,
\end{eqnarray}
which is valid for long bond lengths and small size atoms. 
 % ===================================================================  
 \par
 Since the CO molecule has infinite symmetry around the molecular axis ($C_{\infty v}$ point group),
 the intensity  depends solely on the norm $k$ and polar angle $\theta$ of vector ${\bf k}$,
 i.e. $\left<I({\bf k})\right>_{\varepsilon}=\left<I(k,\theta)\right>_{\varepsilon}$.

 Making use of these additional approximations, we obtain the following formula for the PA-MFPADs within the Muffin-tin plane wave single-scattering approximation,
 \begin{eqnarray}
 \left<I_{single}(k,\theta)\right>_{\varepsilon}
 =
  2 k \alpha \hbar \omega
 \left|
  T_{1}^{\,O} \,
  M_{00}^{1}
  \, \right|^2
 % ----------
 \hspace{4cm}
 \nonumber \\
 % ---------- 
 \times
  \Bigg\{
   1
   +
   \frac{ 2
   \Re \left(
	\,e^{ikR_{}(1-\cos \theta ) }
	f^{\,C}\,(k,\theta) \,
       \right)
   }{\,R_{} } \,
   \cos \theta
   +	 
   \frac{\,\left|f^{\,C}\,(k,\theta)\right|^2 }{\,R_{}^2 } \,
  \Bigg\}
  \label{eq:ave-single-co}
 \end{eqnarray}
where 
$f^{\,C}\,(k,\theta)
 \equiv
 - 4\pi
   \sum_{L}
   T_{l}^{C} \,
   \mathcal{Y}_L(\hat{\bf k}) \,
   \mathcal{Y}_L(\hat{\bf r})
$
is the scattering amplitude
and the superscripts identify the scattering atom, i.e. $C$ for carbon and $O$ for oxygen.
\hatada{The first term corresponds to the direct photoemission process in which the electron does not suffer any subsequent scattering after its excitation from a core state to the continuum state. Because of the polarization averaging, this term brings no angular dependency in the PA-MFPADs.  The third term describes the single scattering by the carbon atom. The second term corresponds to the interference between the direct and singly scattered waves. It is this latter term which creates the so-called \textit{flower shape} at intermediate angles.}

In the case of double-scattering, the PA-MFPADs intensity becomes
\begin{eqnarray} 
% -------------------------------------------------------------------
&& \left<I_{double}(k,\theta)\right>_{\varepsilon}
= % -----------------------------------------------------------------
  2 k \alpha \hbar \omega \,
 \left| \,
		   T_{1}^{\,O}
		   M_{00}^{1}		   
		   \right|^2
		  \nonumber \\
 % ----------  
		  \times &&
		  \Bigg\{ 
		   1
		   +
		   \,
		   \frac{\left|
			  f^{\,C}(k,\theta)
			 \right|^2}
		   {\,R_{}^2\,} \,
  % ----------  
		  +		  
		   \frac{\left|
		   f^{\,O}(k,\pi-\theta)	\,	
		   f^{\,C}(k,\pi)	\,
			 \right|^2}
		   {\,R_{}^4 }\,
 \nonumber \\ %====
 &&         %====
      %====		  
		  + 
		   \frac{
		   2 \Re \left(\,
			  e^{ikR_{}(1-\cos \theta)}\,
			  f^{\,C}(k,\theta) \,
			 \right)
		   }{\,R_{}\,} \,
		   \cos\theta
  \nonumber \\ %====
 &&         %====
      %====		  
		  +
		   \frac{\,
		  2 \Re \left(		   
		   e^{2ikR_{}}\,
		   f^{\,O}(k,\pi-\theta)	\,	
		   f^{\,C}(k,\pi)	\,
		  \right)		   
		   }
		   {\,R_{}^2 }\,		   
		  \cos\theta		  
  \nonumber \\ %====
 &&         %====
      %====
		  + 
		   \frac{
		   2 \Re \left(
		   \,e^{ikR_{}(1+\cos \theta)\,}\,
		   f^{\,*C}(k,\theta) \,
		   f^{\,O}(k,\pi-\theta)	\,	
		   f^{\,C}(k,\pi)	\,
		  \right)		   
		   }{\,R_{}^3 } \,		   
 \Bigg\}.
 \label{eq:ave-double-co}
 \end{eqnarray} 
% =================================================================
\hatada{Compared to the single scattering case, we see the appearance 
of three new terms, the third, fifth and sixth terms. The order of $R$ in the denominator reflects the propagation of the photoelectron, the 
higher this order, the smaller its contribution.
The third term is a purely double scattering intensity term.
\hatadarev{The fifth term corresponds to the interference between the direct and doubly scattered waves, and the sixth term is the single-double scattering interference term.}}
%\yamazaki{\section{Extracting bond length information from the forward- and backward-intensities}}

\subsection{Forward- and Backward-intensities \hatada{as a Function of the Bond Length} : EXAFS-like Formulas} \label{sec:forwardbackward}
 The forward ($\theta=0$)- and backward ($\theta=\pi$)- intensities for single scattering are,
 \begin{eqnarray}  
 &&\left< I_{single}(k,0) \right>_{\varepsilon}
 =
  2 k \alpha \hbar \omega   
 \left| \,
  T_{1}^{\,O} \,
  M_{00}^{1}
  \, \right|^2
  \hspace{1cm}
  \nonumber \\
  &&\times \left\{
   1
   +
   \frac{ 2
   \Re \left(
	f^{\,C}\,(k,0) \,
       \right)	
   }{\,R_{} }
   +	 
   \frac{\,\left|f^{\,C}\,(k,0)\right|^2 }{\,R_{}^2 } \,
  \right\},
  \\
 &&\left< I_{single}(k,\pi) \right>_{\varepsilon}
 =
  2 k \alpha \hbar \omega
 \left| \,
  T_{1}^{\,O} \,
  M_{00}^{1}
  \, \right|^2
    \hspace{1cm}
  \nonumber \\
  &&\times
  \left\{
   1
   -
   \frac{ 2
   \Re \left(
	\,e^{\,2ikR_{}}
	f^{\,C}\,(k,\pi) \,
       \right)
   }{\,R_{} } \,
   +	 
   \frac{\,\left|f^{\,C}\,(k,\pi)\right|^2 }{\,R_{}^2 } \,
  \right\}.  
\end{eqnarray}
 In this case,
 only the backward-intensity $\left< I_{single}  (k,\pi) \right>_{\varepsilon}$ oscillates as a function of $kR_{}$.
% ====================================================================== double scat
 For \hatada{double scattering,} the forward ($\theta=0$) and backward ($\theta=\pi$) intensities become,
% $I^{ave}_{double}({\bf k})$ is reduced to simple forms.
 \begin{eqnarray}  
% ----------------------------------------------------------------------
 &&\left< I_{double}  (k,0) \right> _{\varepsilon}
 = % ----------------------------------------------------------------------
 2 k \alpha \hbar \omega \,
 \left| \,
		   T_{1}^{\,O}
		   M_{00}^{1}		   
		   \right|^2
		  \nonumber \\
 % ----------    
		  \times
		  \Bigg\{
		   1
		  &&+
		   \,
		   \frac{\left|
			  f^{\,C}(k,0)
			 \right|^2}
		   {\,R_{}^2\,} \,
  % ----------  
		  +		  
		   \frac{\left|
		   f^{\,O}(k,\pi)	\,	
		   f^{\,C}(k,\pi)	\,
			 \right|^2}
		   {\,R_{}^4 }\,
  \nonumber \\
          %====
          %====
          %====		  
   		   % &
                  && + 
   		   \frac{
   		   2 \Re \left(\,
   			  f^{\,C}(k,0) \,
   			 \right)
   		   }{\,R_{}\,} \,
   % \nonumber \\ %====
 % &         %====
 % \left.     %====		  
		  +
		   \frac{\,
		  2 \Re \left(		   
		   e^{2ikR_{}}\,
		   f^{\,O}(k,\pi)	\,	
		   f^{\,C}(k,\pi)	\,
		  \right)		   
		   }
		   {\,R_{}^2 }\,		   
 % \right.
  \nonumber \\ %====
 % &         %====
 % \left.     %====
		  &&+ 
		   \frac{
		   2 \Re \left(
		   \,e^{2ikR_{}\,}\,
		   f^{\,*C}(k,0) \,
		   f^{\,O}(k,\pi)	\,	
		   f^{\,C}(k,\pi)	\,
		  \right)		   
		   }{\,R_{}^3 }		   
 \Bigg\} \label{eq:double_zero}  ,
 %  \nonumber
  \\
% % -------------------- 
&& \left<I_{double}(k,\pi)\right>_{\varepsilon}
  = % ----------------------------------------------------------------------
 2 k \alpha \hbar \omega \,
 \left| \,
		   T_{1}^{\,O}
		   M_{00}^{1}		   
		   \right|^2
		  \nonumber \\
 % ----------  
		  \times
		  \Bigg\{
		   1
		  &&+
		   \,
		   \frac{\left|
			  f^{\,C}(k,\pi)
			 \right|^2}
		   {\,R_{}^2\,} \,
  % ----------  
		  +		  
		   \frac{\left|
		   f^{\,O}(k,0)	\,	
		   f^{\,C}(k,\pi)	\,
			 \right|^2}
		   {\,R_{}^4 }\,
%
%  % \right.
  \nonumber \\ %====
%  % &         %====
%  % \left.     %====		  
		  && - 
		   \frac{
		   2 \Re \left(\,
			  e^{2ikR_{}}\,
			  f^{\,C}(k,\pi) \,
			 \right)
		   }{\,R_{}\,} \,
% %
%  \right. \nonumber \\ %====
%  &         %====
%  \left.     %====		  
		  -
		   \frac{\,
		  2 \Re \left(		   
		   e^{2ikR_{}}\,
		   f^{\,O}(k,0)	\,	
		   f^{\,C}(k,\pi)	\,
		  \right)		   
		   }
		   {\,R_{}^2 }\,		   
%  % \right.
  \nonumber \\ %====
%  % &         %====
%  % \left.     %====
		 &&+ 
		   \frac{
		   2 \Re \left(
		   f^{\,*C}(k,\pi) \,
		   f^{\,O}(k,0)	\,	
		   f^{\,C}(k,\pi)	\,
		  \right)		   
		   }{\,R_{}^3 }		\Bigg\}  
  . \label{eq:double_pi}
 \end{eqnarray}
 Now, the forward-intensity also oscillates with \ota{$2kR$}.
  %		  
 %\Bigg)
These equations \hatada{may} explain the trends of the PA-MFPADs.
%For Muffin-Tin calculation the forward intensities oscillates  two or more scatterings are included while the backward intensities oscillate in all calculations as shown in figure~\ref{fig:oscillation}.

\begin{figure}[htb]
\includegraphics[width=1.0\linewidth]{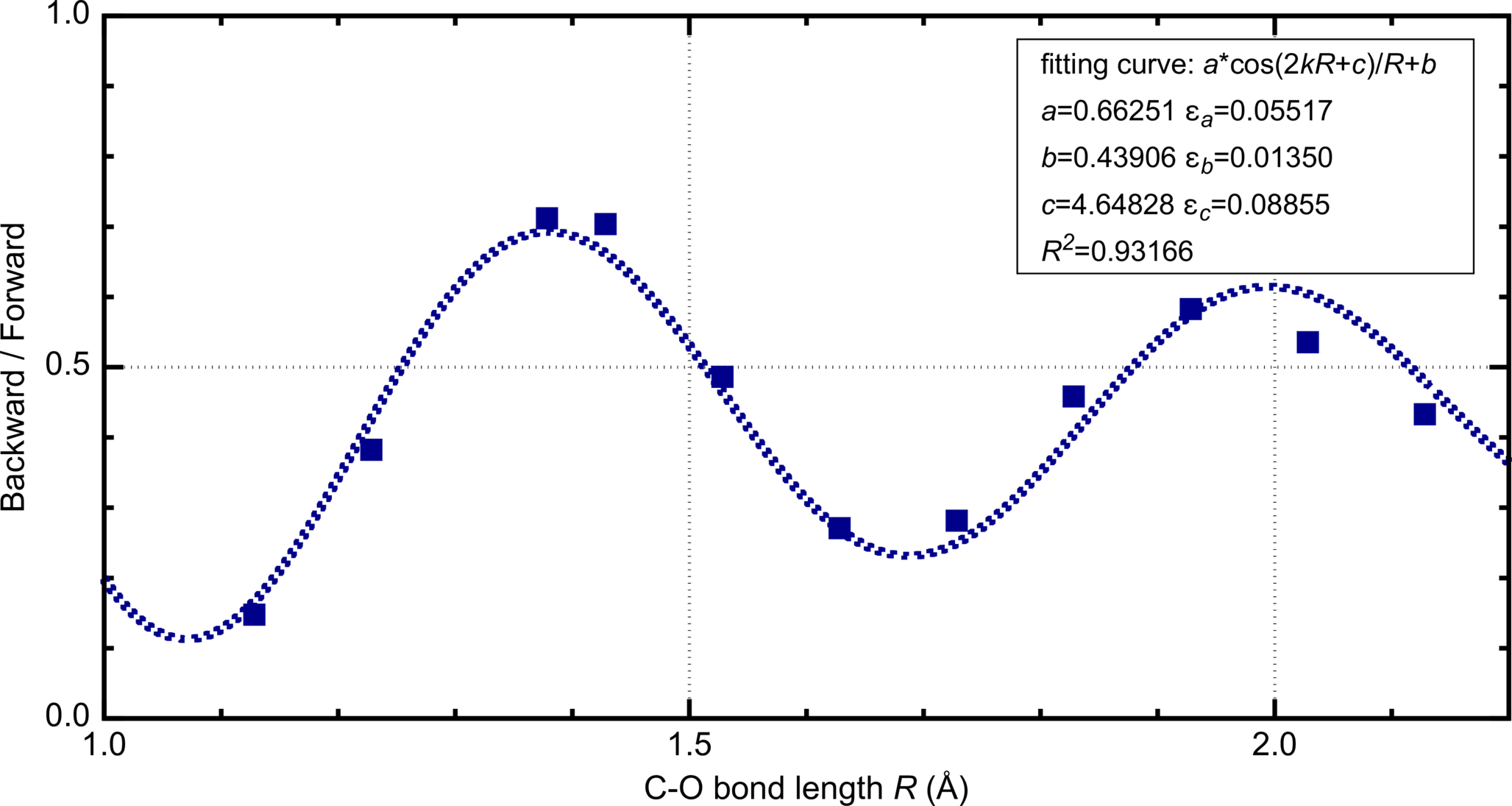}
\caption{
\label{fig:oscillation}
\yamazaki{\hatada{Ratio of the backward-intensity to the  forward-intensity} of the oxygen $1s$ PA-MFPADs of CO${}^{2+}$ as a function of the C-O bond length $R$ from 1.1283 \r{A}\, (equilibrium bond length in the ground state) to 2.1283 \r{A}\ for the 1$\sigma^{-1} 5\sigma^{-2}$ state,  calculated with the Full-potential multiple scattering method (Blue dot) and compared to our least square fitted result using equation~\ref{eq:fb_ratio} (dashed line).} 
}
\end{figure}

From an experimental point of view, it is significantly easier to extract the variation of the ratio of the backward- and forward-intensities rather than the intensity variations of the backward- and forward-intensities themselves. So, we  consider \ota{their} ratio now. Figure~\ref{fig:oscillation} depicts the intensity ratio of the forward- and backward-intensity peaks of the oxygen $1s$ PA-MFPADs of CO${}^{2+}$ as a function of the C-O bond length $R$ from 1.1283 \r{A} (equilibrium bond length in the ground state) to 2.1283 \r{A}. They were obtained from our Full-potential multiple scattering calculations~\cite{Ota2020a}. We see clearly here the EXAFS-like  oscillations in the ratio. \hatada{In order to} model these oscillations {with an analytical formula}, we \hatada{employ} the single-scattering approximation\ota{, which} is known to capture the EXAFS oscillations: 
\begin{eqnarray}
\hatada{\eta(k,R)}&\equiv& \hatada{\frac{\left<I_{single}( k, \pi)\right>_{\varepsilon}}
{\left<I_{single}( k, 0)\right>_{\varepsilon}}}
\nonumber \\
&=&   \yamazaki{\frac{a(k,R)}{R} \cos (2 k R-\hatada{\phi^C} (k,\pi))+b(k,R)} \label{eq:fb_ratio}, 
\end{eqnarray} 
\hatada{where}
\yamazaki{\begin{eqnarray} 
a(k,R) &\equiv -\frac{2 \hatada{R^2}\left|f^{C}(k, \pi)\right|}{\hatada{R^2}+2\hatada{R}~\Re\left(f^{C}(k,0)\right)
+\hatada{\left|f^{C}(k,0)\right|^{2}}}, \\ \label{eq:fb_ratio_a}
%&\sim 
%-\frac{2 \left|f^{C}(k, \pi)\right|}{R} \\
%
b(k, R)  &\equiv  \frac{R^{2}+\left|f^{C}(k,\pi)\right|^{2}}{R^{2}
+2 R~\Re\left(f^{C}(k,0)\right)
+\left|f^{C}(k,0)\right|^{2}}. \label{eq:fb_ratio_b}
%&\sim 1 \\
\end{eqnarray}}
\hatada{
The ratio of the backward-intensity against the forward-intensity oscillates with a frequency $2kR$. The function $\phi(k,\pi)$ is the  phase factor of the back scattering amplitude from the carbon atom, $f^{C}(k,\pi)=\left|f^{C}(k,\pi)\right|\exp{[i\phi^C(k,\pi)]}$.} 
%In figure~\ref{fig:oscillation} we show the least square fitting with this equation to the calculated data done by Full-potential multiple scattering theory.}%as s}
%\ota{
%We can now see explicitly that, in the framework within the Muffin-tin approximation and single-scattering, 
%the ratio of backward-intensity against forward-intensity oscillate with frequency $2kR$. }%as shown in figure~\ref{fig:oscillation}. }

%\yamazaki{The oscillatory structure in the backward/forward-scattering intensity ratio  is clearly seen in the Full-potential multiple scattering. Its oscillation is dominated by the single backward scattering as discussed in the preceeding paper [ref] . 
%The frequency of the oscillation is $2kR$, as expected in equation~\ref{eq:fb_ratio}. 
%KU: I do not think this sentense is correct. Could you confirm or deconfirm it!}} 
%not exactly $2kR$ since 
%the scattering amplitude $f$ also depends on the bond length, but from the formulation it is clear that the oscillations are caused by the interference between waves with different scattering numbers, and the frequency of the oscillation depends on the energy and coupling length. 

%\yamazaki{\subsection{Forward- backward- intensity ratio}}

%\yamazaki{The bond length dependency of the Forward- backward- intensity ratio $\rho(k,R) \equiv I&^{ave}_{double}(k,\pi) / I&^{ave}_{double}(k,0)$ is similar both in the Full-potential  and second order Muffin-tin calculations as shown in Fig XX. This indicates that one can estimate $R$ from an simplified expression of $\rho(k,R)$ based on second order Muffin-tin calculation.  \textbf{Fukiko, Please write the brief derivation of simplified $\rho(k,R)$ here}.}

%\ueda{I do not think we need this subsection separately. It can be either in the previous subsection or in III A. But we should wait for the feedback from Toyama group. We still do not know what we can do.}\\

%\yamazaki{\section{Extracting bond length information from interference structure}}

\ueda{\section{Extraction of the Bond Length Information from the Forward-Backward Peaks in the PA-MFPADs}}

In this section, we propose a new experimental procedure to determine the time evolution of the bond length of diatomic molecules. This procedure is  based on the approximation of  \hatada{$\eta(k,R)$} given by equation~\hatada{\ref{eq:fb_ratio}}. % the Young's formula \ref{Young}. 

We begin by noting that $a(k,R)$, $b(k,R)$ and $\phi^C(k,\pi)$ vary slowly as a function of $R$ in equation ~\ref{eq:fb_ratio}. Therefore, we may consider them as constant. As a consequence, the \hatada{forward-backward intensity} ratio \hatada{$\eta(k,R)$}  can be adequately fitted by the EXAFS-like equation 
\begin{eqnarray}
    \hatada{\eta}(k,R)  =  \frac{a}{R} \cos (2 k R-\phi))+b . \label{eq:exafs-def}
\end{eqnarray}
In figure~\ref{fig:oscillation}\ota{,} we show the comparison between this equation with a least square fitting and  the theoretical data calculated with our Full-potential multiple scattering theory.
Here the correlation coefficient is $r^2 = 0.93166$, suggesting that our fitting of $a$, $b$ and $\phi$ is reasonable.
%(figure~\hatada{2}). 
This means that our equation of \hatada{$\eta(k,R)$} works as \didier{an} experimental  "bond length ruler".

\begin{figure}[htb]
\includegraphics[width=0.8\linewidth]{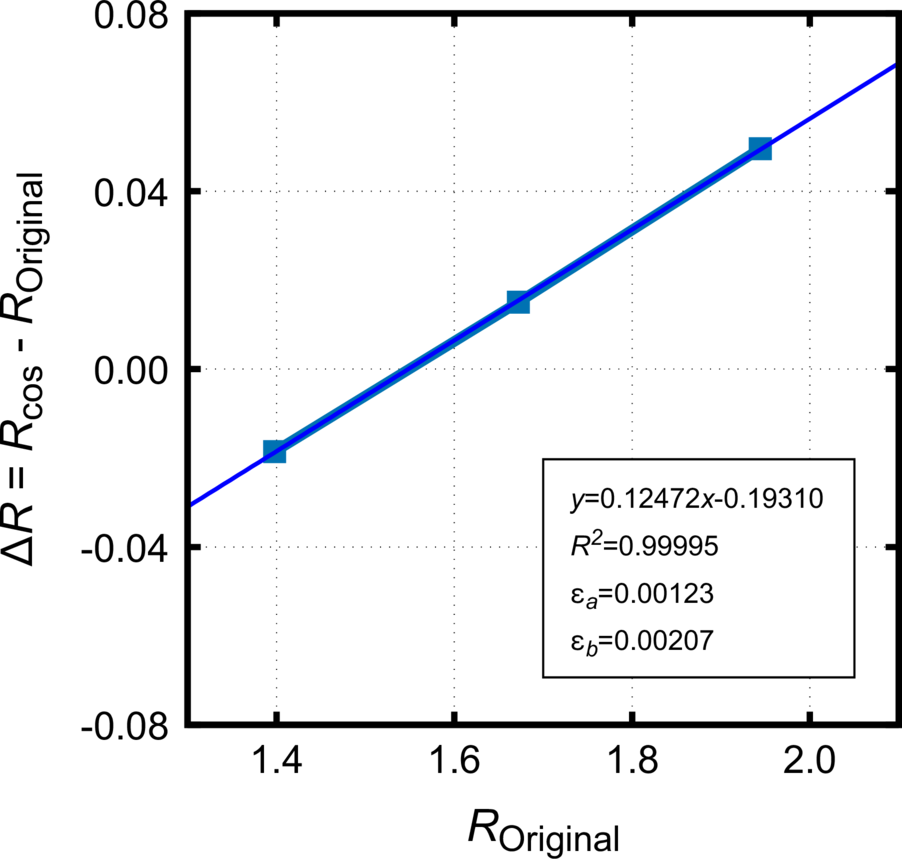}
\caption{ The deviation of the C-O bond length evaluated from the fitting of equation~\ref{eq:fb_ratio} against \hatada{the} Full-potential calculations displayed in figure~\ref{fig:oscillation}.
 }\label{fig:diff_R}
\end{figure}

In figure~\ota{\ref{fig:diff_R}}, we have plotted the deviation of the C-O bond length \ueda{at} the peak positions of \hatada{$\eta(k,R)$} against that of our Full-potential calculations obtained by numerical differentiation.   
The accuracy of the extracted bond length $R$ employing equation~\ref{eq:exafs-def} is estimated to be better than 0.1 \r{A} as shown in figure~\ref{fig:diff_R} (root mean square error: 0.032 \AA).
The  error on the C-O bond length $\Delta R$ is a linear function of the \ota{original} C-O bond length $R_\mathrm{Original}$: $\Delta R = 0.12472 R_\mathrm{Original} -1.9310$ (correlation coefficient $r^2 = 1.00$). The upper bound of the relative error $\Delta R/R_{\mathrm{Original}}$ is estimated to be 12.5 \% in the limit of sufficiently large C-O \ota{bond lengths}.  

 The major origin of the fitting error in $R$ comes for the neglect of the variations of $a, \phi,$ and $b$ (see  equations \ref{eq:fb_ratio}-\ref{eq:fb_ratio_b}). We have also neglected the chemical shift of $k$, which can be  estimated to 2-3 \% 
from our RASPT2/ANO-RCC-VQZP results as detailed in the preceding paper~\cite{Ota2020a}. 
In addition, equation~\ref{eq:fb_ratio} does not contain the multiple scattering  (second order terms in \otarev{ equations~\ref{eq:double_zero} and~\ref{eq:double_pi}}, and other higher order terms) included in our Full-potential multiple scattering calculations. These additional terms will modulate the phase of the EXAFS-like fitting in equation~\ref{eq:exafs-def} and explain the small error induced by our ruler formula on the C-O bond length. 

\begin{figure}[htb]
\includegraphics[width=\linewidth]{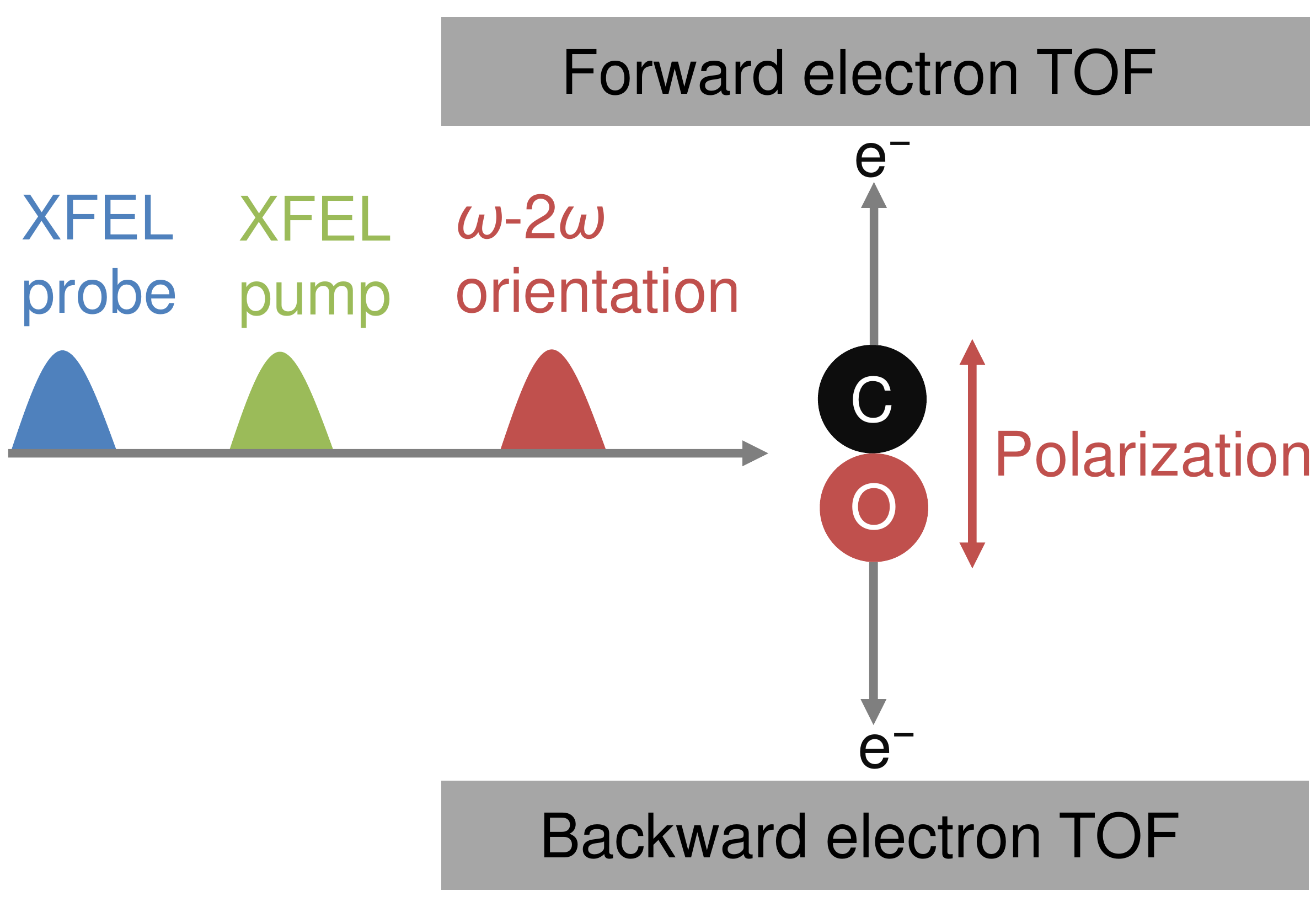}
\caption{A schematic description of an %\yamazaki{alternative}
experimental set-up for the forward-/backward-intensity ratio %\hatada{$\eta(k,R)$} 
measurements \yamazaki{without COLTRIMS--Reaction Microscope}. Here the forward-/backward-intensities are measured for molecules spatially oriented impulsively by the phase-coherent two-color (e.g., 800 nm and 400 nm) double pulses, in the direction parallel to the molecular orientation that coincides with the linear polarization axis of the two XFEL pulses. Although polarization average is not taken in this scheme, the ratio thus obtained coincides with $\eta(k,R)$ defined in equation~\ref{eq:fb_ratio}. 
%The ratio measured in this scheme also exhibits the oscillatory structure similar to the one in \fref{fig:oscillation} described by equation 20 with the same frequency $kR$.             
\label{fig:fb_detector}  
}
\end{figure}

%Experimentally, one can extract can  extract η(k,R)  from  PA-MFPADs recorded  by  COLTRIMS--Reaction  Microscope. 
%Below, we describe another detection scheme combiningan impulsive orientation of the molecule [28]. 
\ueda{Experimentally, 
%one can \hatada{measure} %extract
%\hatada{$\eta(k,R)$} from PA-MFPADs recorded 
\ota{one can measure the ratio $\eta(k,R)$ between the forward- and backward-intensities of the PA-MFPADs}
with a COLTRIMS--Reaction
Microscope. Below, we describe other detection schemes combining impulsive orientation of the molecule \yamazaki{~\cite{Koch2019}}. The orientation of linear molecules that have dipole moments may be prepared by the phase controlled two \yamazaki{$\omega-2\omega$} pulses, say, at 800 nm and 400 nm \yamazaki{\cite{Wu2010}}.  Orientating CO molecules in space, one can measure \hatada{$\eta(k,R)$} with two electron time-of-flight (TOF) spectrometers placed face to face in the direction of the axis of parallel polarization of four sequential pulses, as shown in figure~\ref{fig:fb_detector}. This tandem TOF apparatus is often used for the carrier envelope phase measurements of femotosecond/attosecond optical pulses and referred to as  stereo-ATI~\cite{Paulus2001}. Alternatively, one can use a velocity map imaging \yamazaki{(VMI)}~\cite{Minemoto2016} electron spectrometer instead of two TOF spectrometers. These approaches may also work for low repetition-rate XFELs~\cite{Minemoto2016} or high-power high-harmonic generation light sources~\cite{Fu2020}  as long as two color X-ray pulses are available. For these forward-backward photoemission measurements of molecules with the maximally oriented rotational wave packet, the jitters between the optical lasers and pump and probe X-ray pulses may need to be corrected~\cite{Katayama2016,Kumagai2018,Rolles2018} by the shot-by-shot monitoring technique, which has already been implemented in the XFEL facilities. }
%The  photoelectrons ejected to the forward and backward directions are detected as a function of delay time $t$ using the two individual TOF detectors or velocity map imaging detector. }
\\

% ======================================================================
% ======================================================================

\section{Conclusions}
In this work, we have proposed a new theoretical method to extract from the experimental data the dissociation dynamics of diatomic molecules via time-resolved PA-MFPADs measurements using the COLTRIMS--Reaction Microscope and a two-color XFEL pump-probe set-up.
The Muffin-tin \hatada{single scattering approximation} leads to the  \yamazaki{simple EXAFS-like formula \ota{of} equation~\ref{eq:fb_ratio}}. 
This formula acts as an experimental "bond length ruler" controlled by only three parameters related to experimental results.
\hatadarev{
In this study we have only considered the case of a two atom molecule. However, our method is also valid for polyatomic molecules where EXAFS-like oscillations are also present. At lower energies this simple single scattering picture will break down and multiple scattering need be applied.}

\yamazaki{The accuracy and applicability of equation~\ref{eq:fb_ratio} has been  numerically checked against our {\it in-silico} reference bond length dependent PA-MFPADs of the CO\ota{$^{2+}$} molecule detailed  \yamazaki{in the preceding paper~\cite{Ota2020a}}. We have demonstrated analytically that the EXAFS-like formula gives us the bond length of the dissociating CO\ota{$^{2+}$} molecule with an accuracy of \yamazaki{0.05-0.1} \r{A}. }

\ueda{This method will become applicable to experiments with the COLTRIMS--Reaction Microscope end station~\cite{Kastirke2020} installed in the SQS beam line of European XFEL~\cite{Tschentscher2017} and the DREAM end station\hatadarev{~\cite{LCLSII}} equivalent to the COLTRIMS-Reaction Microscope
%[https://lcls.slac.stanford.edu/instruments/l2si]
installed in LCLS-II~\cite{Duris2020}, when the two-color femtosecond soft XFEL starts operating for time-resolved studies in the near future~\cite{Duris2020, Serkez2020}.} \ueda{XFEL facilities usually have a femtosecond NIR laser systems that may be synchronized with XFEL pulses and thus used for orientating target molecules. Then, measurements for the forward-backward intensity ratios are straightforward and variations of the bond-length can be extracted from such measurements.}  
We hope that time-resolved PA-MFPADs will become a new attractive pathway to make molecular movies visualizing intramolecular reaction.

%but the value o indicates the deviation of experimental condition from the high-energy classical limit.

%\comment{ In other words, for neighboring peaks around the direction $\theta=\pi/2$, we should choose $\alpha$ other than $2$ 
%and its estimation is an subject for the future study. }
%\\
%FIG11(I'll put the figure later)\\
%Figure 11 shows that the nuclear distance calculated by equation~\ref{Young} from the peak position of PA-MFPADs %obtaind by multiple scattering calculation within Full-potential method.\\
%.....
% =================================================================
\section*{Acknowledgement}
\hatadarev{We would like to acknowledge C. R. Natoli for fruitful discussions.}
This work was performed under the Cooperative Research Program of ``Network Joint Research Center for Materials and Devices''. K.H. acknowledges funding by JSPS KAKENHI under Grant No. 18K05027 and 17K04980. K. Y. is grateful for the financial support from Building of Consortia for the Development of Human Resources in Science and Technology, MEXT, and JSPS KAKENHI Grant Number 19H05628. 
\section*{References}
\bibliographystyle{unsrt}
\bibliography{fukiko}
\end{document}